%% file: ICRC2025_IceCube.tex
\documentclass[a4paper,11pt]{article}

\usepackage{pos}
\usepackage{lipsum} 
\usepackage{subcaption}
\usepackage{caption}
\usepackage{lineno}

\title{Astrophysical neutrino flux measurement and search for tau neutrino induced cascades with 11 years of IceCube data}

\ShortTitle{Astrophysical flux and flavor measurements}

\author{The IceCube Collaboration \\{\normalsize \normalfont(a complete list of authors can be found at the end of the proceedings)}\\}

\emailAdd{zheyang.chen@stonybrook.edu}
\emailAdd{zelong.zhang.1@stonybrook.edu}
\emailAdd{joanna.kiryluk@stonybrook.edu}

\abstract{

IceCube measured the diffuse astrophysical neutrino flux for all flavors up to PeV energies. The high energy (TeV-PeV) IceCube cascade sample is particularly effective at selecting electron and tau neutrinos. We present the results of Single Power Law (SPL) and Broken Power Law (BPL) flux measurements based on 11 years of cascade data. From this cascade sample, we study the identification of high energy (\textasciitilde100 TeV) tau neutrinos by detecting a double cascade signature produced by a charged-current neutrino interaction and the subsequent decay of the tau lepton. A Boosted Decision Tree (BDT) is employed to search for the double cascade signature, achieving a significantly improved signal-to-background ratio of 9:1 compared to previous analyses. The selected sample has a tau-neutrino purity of approximately 90\% and a weighted mean reconstruction error on the tau decay length of about 4m. We present sensitivities for a maximum likelihood fit of the flavor composition and constraints on the astrophysical neutrino flavor ratios using this sample in combination with IceCube’s northern muon neutrino-induced track sample.

\vspace{4mm}

{\bfseries Corresponding authors:}
Zheyang Chen$^{1*}$, 
Zelong Zhang$^{1}$, 
Joanna Kiryluk$^{1}$, 
\\
{$^{1}$ \itshape Stony Brook University}\\
[4mm]
$^*$ Presenter
}

\input{ICRCdetails.tex}

\begin{document}

\maketitle

\section{Introduction}\label{sec1}

The discovery of a diffuse flux of high-energy astrophysical neutrinos by the IceCube Neutrino Observatory \cite{doi:10.1126/science.1242856} has opened a new era in multi-messenger astrophysics. These neutrinos are believed to be produced in hadronic interactions within powerful cosmic accelerators such as active galactic nuclei or supernovae. Measuring the properties of this flux — including its energy spectrum and flavor composition — provides critical insight into the origin of high-energy cosmic rays and probes neutrino physics over cosmological distances. Therefore, several analyses have been performed in various channels within IceCube collaboration to measure both the astrophysical neutrino energy spectrum \cite{PhysRevD.104.022002, PhysRevD.91.022001, vedant_thesis, Abbasi_2022, Abbasi:2023p4, PhysRevLett.125.121104, Naab:2023xcz} and the flavor composition \cite{Usner2018Search,juliana_2022,PRL_doublepulse,Lad:2023u9,Basu:2023i3}.

The IceCube detector is a cubic-kilometer neutrino observatory located at the South Pole, instrumented with 5,160 digital optical modules (DOMs) deployed along 86 vertical strings. It detects neutrinos by observing Cherenkov light from charged secondary particles produced in neutrino interactions with the Antarctic ice. Events are categorized as either track-like or cascade-like based on their topology. Track events arise primarily from muons produced in $\nu_\mu$ charged-current (CC) interactions, and provide excellent angular resolution of \textasciitilde1.5 degree \cite{PhysRevD.104.022002,Abbasi:2023p4}. In contrast, cascade events originate from CC interactions of $\nu_e$ and $\nu_\tau$ along with neutral-current (NC) interactions of all flavors. Cascades yield a largely flavor-inclusive signature and achieve an excellent energy resolution of about 7\% at around 100 TeV. The two topologies are complementary in reconstructing the full astrophysical neutrino flux.

In this work, we present an updated measurement of the astrophysical electron and tau neutrino flux using 11 years of cascade data from IceCube, including fits to both single and broken power-law models. We further introduce a method for identifying tau neutrino events using the double cascade signature of $\nu_\tau$ CC interactions. The resulting high-purity $\nu_\tau$ sample enables enhanced sensitivity to the flavor composition of diffuse neutrino flux. We evaluate expected flavor constraints by combining this tau-enriched cascade sample with IceCube’s northern track sample \cite{Abbasi_2022}, demonstrating the power of a joint analysis across event types.

\section{Astrophysical Neutrino Flux Measurement}\label{sec2}
In this work we extend the 6-year IceCube cascade analysis \cite{PhysRevLett.125.121104} to an 11-year dataset (2010–2020), update the ice model \cite{Rongen:771097, Abbasi_2024} used for event simulation and reconstruction, and refine the treatment of systematic uncertainties. Considering both single and broken power law parameterizations of the astrophysical neutrino flux, we adopt the same binning scheme and fitting framework as \cite{PhysRevLett.125.121104}, and perform a maximum-likelihood fit with an updated effective likelihood\cite{Arguelles:2019izp} that propagates the statistical uncertainty of the Monte Carlo (MC) datasets.  This procedure yields the most up-to-date measurement of the diffuse astrophysical neutrino flux in the cascade channel, with markedly reduced uncertainties, as shown in Sec. \ref{subsec:astrophysical flux measurement}.

After recalibration of the DOM efficiencies by the IceCube Collaboration, the datasets used here were reprocessed, yielding the ``pass2'' datasets \cite{aartsen2020situ}. We verified, during unblinding, that the original selection cuts and Boosted Decision Tree (BDT) classifier from 6-year Cascade analysis still delivers excellent data–MC agreement: a binned \(\chi^{2}\) goodness-of-fit test yields a p-value of \(0.50\), indicating no significant discrepancy. This selection is applied to the full 11-year data and classifies data into three topological classes—cascades, starting-tracks, and through-going tracks. Cascades constitute the astrophysical-neutrino signal sample, starting-tracks provide the atmospheric-neutrino background control sample, and through-going tracks supply the atmospheric-muon background control sample.

With this selection, the number of events in cascade sample increases from roughly \(4{,}800\) in the 6-year study to more than \(14{,}000\).\footnote{The apparent mismatch between the livetime scaling ($11/6\!\simeq\!1.8$) and the event–count scaling ($14\,000/4\,800\!\simeq\!2.9$) comes from the way season 2010 and 2011 were handled in the 6-year analysis. That study retained only 113 very-high-energy cascades from those two years. If we compare events rate of this analysis (14000/11) and events rate of season 2012 to season 2015 (4700/4), they are consistent to each other.} The Combined Fit analysis \cite{Naab:2023xcz} incorporates the cascade and starting-track samples presented here, but substitutes our track sample with a Northern-sky track sample \cite{Abbasi_2022}, yielding tighter control of the atmospheric-muon background. In addition, the two studies rely on independent Monte-Carlo simulations and distinct fitting frameworks.

The modeling of ice optical properties, such as scattering lengths, absorption lengths, and hole-ice properties, have been refined \cite{Rongen:771097, Abbasi_2024}. Using the updated \textsc{SPICE-3.2.1} model, angular resolution for high-energy cascades (\(E_{\text{cascade}}>100~\text{TeV}\)) improves to about \(5^{\circ}\). A unified, model-independent parameterization of the hole-ice properties \cite{Eller:2023ko}, governed by two parameters \(p_{0}\) and \(p_{1}\), now affords a consistent treatment of uncertainties associated with the refrozen columns surrounding DOM.

Atmospheric neutrinos are often accompanied by atmospheric muons that can trigger the veto system—a \emph{self-veto} that biases a neutrino-only simulation toward higher rates. While the 6-year Cascade analysis used an earlier version of self-veto passing rate calculator \cite{PhysRevD.90.023009} for a per-event event weight correction, an updated version is now implemented which (i) derives parent spectra from cosmic-ray interactions, (ii) incorporates stochastic muon-energy losses, and (iii) enables systematic-uncertainty studies of fractions of atmospheric neutrinos passing the selections \cite{Argüelles_2018}. 

Finally, to propagate MC statistical uncertainties in each analysis bin we replace the usual Poisson likelihood with an effective likelihood \cite{Arguelles:2019izp}. According an Asimov profile log-likelihood scan, this choice broadens the profile log-likelihood contours slightly: the \(1\sigma\) uncertainties on the spectral index \(\gamma_{Astro}\) and flux normalization \(\Phi_{0}\) of astrophysical neutrino flux each increase by about \(5\%\), while the best-fit values remain unchanged.

\section{Astrophysical Flavor Measurement}\label{sec3}
The flavor composition of the astrophysical neutrino flux is a fundamental observable in neutrino astronomy. While neutrinos are produced in flavor-specific processes at their sources, propagation over cosmological distances leads to mixing through neutrino oscillations. For standard neutrino production via pion decay in sources, an initial flavor ratio of $(\nu_e : \nu_\mu : \nu_\tau) = (1 : 2 : 0)$ is expected to evolve to approximately $(1 : 1 : 1)$ at Earth and assumed in Section~\ref{sec2}. Measuring this flavor composition provides a test of standard oscillation physics and serves as a probe of the underlying neutrino production mechanisms.

In IceCube, flavor separation is challenging because multiple neutrino flavors can produce similar topologies. Muon neutrinos are efficiently identified by their long track-like signatures, while electron and tau neutrinos typically yield cascade-like events that are difficult to distinguish. However, tau neutrinos interacting via charged-current processes can exhibit a distinct double cascade signature: one cascade at the interaction vertex and another from the tau decay. This topology becomes resolvable when the tau decay length exceeds $\sim$10 meters, typically at energies above several tens of TeV. The decay length is strongly correlated with the tau energy via $L_\tau \approx 49.5~\text{m} \times (E_\tau / 1~\text{PeV})$, making spatial separation a powerful handle for identifying $\nu_\tau$ events.

\subsection{Double Cascade Selection}

The cascade dataset provides a sample largely free of penetrating muon tracks, making it well-suited for identifying double-cascade topologies from $\nu_\tau$ CC interactions, where a tau lepton is produced and promptly decays. About 83\% of tau decays yield hadronic or electronic final states that can form a visible second cascade; the remaining $\sim$17\% are muonic and exhibit track-like topologies not targeted in this analysis. The dominant background consists of cascade-like events from $\nu_e$ CC and NC interactions of all flavors, as well as starting-track events from $\nu_\mu$ CC interactions.

To identify double cascade $\nu_\tau$ events, we follow a reconstruction-based approach similar to previous IceCube analyses~ \cite{Usner2018Search,juliana_2022}, using the \texttt{Taupede} algorithm~ \cite{Hallen:2013}, which fits the energies and positions of two spatially separated cascades. This analysis uses updated simulations and reconstructions incorporating the latest ice modeling results~ \cite{Abbasi:2023fT}. Two key observables—tau decay length and energy asymmetry defined as: $A = (E_1 - E_2)/(E_1 + E_2)$—provide strong discrimination between true double cascade events and backgrounds. A set of initial selection cuts ("precuts") are applied to suppress low-energy and poorly resolved events while preserving signal efficiency. These include a minimum reconstructed energy of $10^{4.5}$GeV, a decay length threshold of at least 10 meters, and containment conditions that require the interaction and decay vertices to be within the instrumented volume and outside the dust layer region\footnote{The \textit{dust layer} refers to a 100\,m thick region of glacial ice located at a depth of approximately 2000\,m in the IceCube detector, characterized by an enhanced concentration of dust, which reduces optical clarity and degrades photon propagation, affecting event reconstruction quality.}. Following the precuts, two Boosted Decision Trees (BDTs) are used to enhance signal purity: the first targets single-cascade backgrounds (e.g., $\nu_e$ CC and NC events), and the second rejects starting-track backgrounds from $\nu_\mu$ CC interactions. The BDTs are trained on variables from both single and double cascade reconstructions, with the most important inputs including the log-likelihoods under each hypothesis, decay length, energy asymmetry, individual cascade energies, and total recorded charge.

\begin{figure}[htbp]
  \centering
  \begin{subcaptionbox}{Event rates for each neutrino flavor after successive selection steps in one year of simulation.The x-axis labels indicate the cut levels, where Energy cut and Length cut are reconstructed energy and tau decay length cut that are part of precuts. We achieve a signal to background ratio of about 9 to 1 after all cuts.\label{fig:double_cascade_selection_eventrate}}[0.45\textwidth]
    {\includegraphics[width=\linewidth]{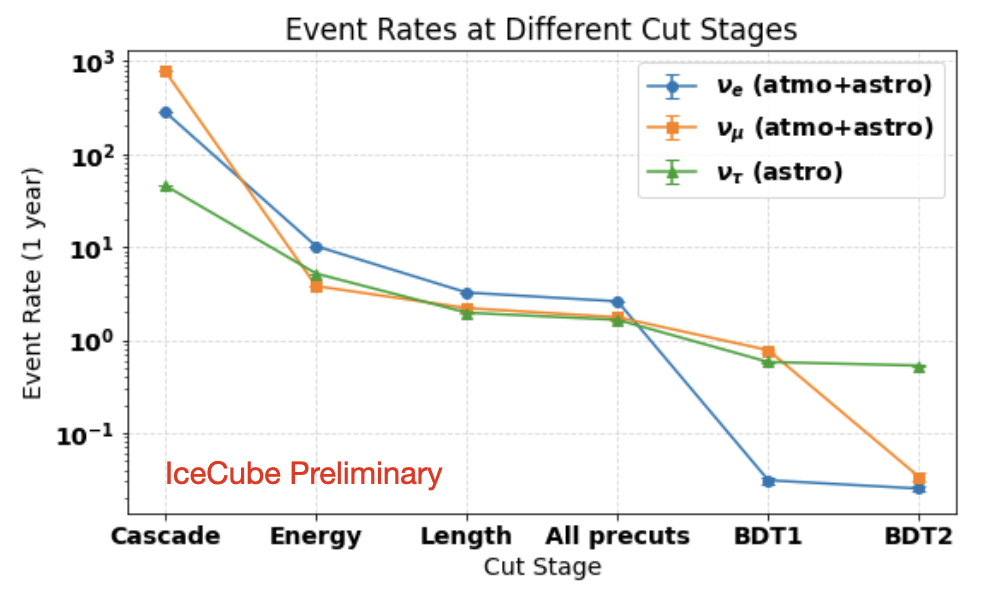}}
  \end{subcaptionbox}
  \hfill
  \begin{subcaptionbox}{Selection efficiency and purity ratio for long double cascade ($L_\tau > 10$~m) $\nu_\tau$ events at each cut level. Efficiency is defined as the fraction of long double cascade events surviving relative to the initial sample, and the ratio is the fraction of such events among all events at each level.\label{fig:double_cascade_selection_efficiencypurity}}[0.45\textwidth]
    {\includegraphics[width=\linewidth]{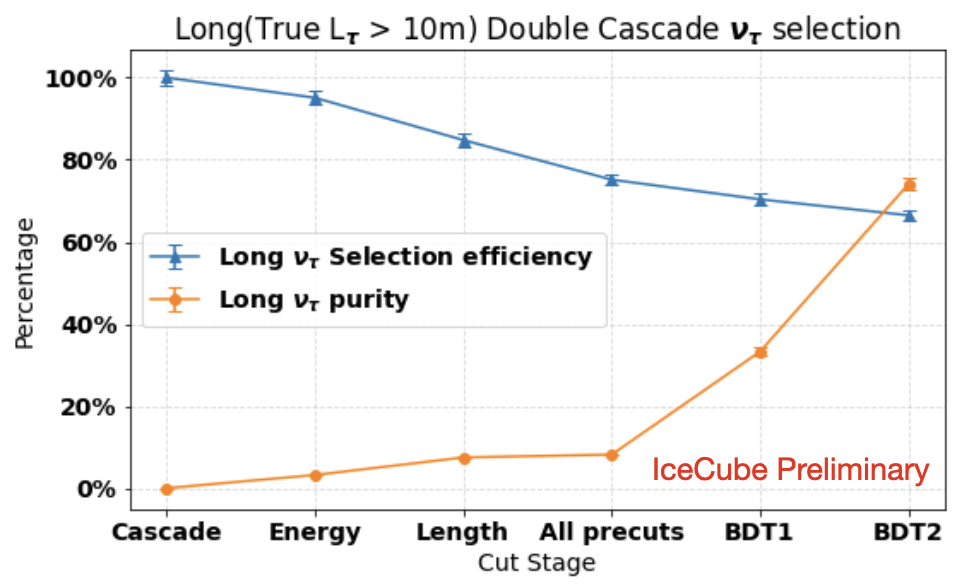}}
  \end{subcaptionbox}
  \caption{Backgrounds from $\nu_e$ and $\nu_\mu$ are suppressed while maintaining a high-purity $\nu_\tau$ sample.}
  \label{fig:double_cascade_selection}
  \vspace{-10pt}
\end{figure}

After applying the precuts and optimized BDT selections, we obtain a $\nu_\tau$-enriched double cascade sample with an estimated event rate of approximately 0.6 per year. The sample has a $\nu_\tau$ purity of about 90\%, defined as the fraction of selected events originating from true $\nu_\tau$ interactions. Among these, 99\% exhibit a double cascade topology and 82\% have a true tau decay length exceeding 10 meters as shown in Figure~\ref{fig:double_cascade_selection_efficiencypurity}. The weighted mean reconstruction error on the tau decay length for selected $\nu_\tau$ events is approximately 4 meters.

This analysis builds on and improves on previous tau neutrino searches in IceCube. Compared to earlier straight-cut based analyses like High-Energy Starting Events (HESE) \cite{juliana_2022,Lad:2023u9} and Medium Energy Starting Events (MESE) \cite{Basu:2023i3}, which applied fixed thresholds on reconstructed variables, our BDT-based selection achieves a higher signal-to-background ratio and selection efficiency, yielding improved $\nu_\tau$ purity over the ~70\% reported in MESE. In contrast to the convolutional neural network (CNN)-based double pulse analysis \cite{PRL_doublepulse}, which does not provide a reconstructed event topology or a flavor composition measurement, our approach yields a similarly high-purity sample with the added benefit of full topological reconstruction. These improvements make the selected sample suitable for use in likelihood-based flavor composition analyses.

\subsection{Analysis Method}
The complete cascade data set is divided into two exclusive subsets: the $\nu_\tau$-enriched sample that passes the double cascade selection and the remaining single cascade events. This separation ensures no overlap between the two cascade-based inputs. To evaluate sensitivity to the astrophysical neutrino flavor composition, we perform a binned likelihood analysis combining three complementary event samples: the 11 years $\nu_\tau$ enriched double cascade sample, the 11 years single cascade sample, and a 9.5 years northern sky track sample \cite{Abbasi_2022_numu} dominated by $\nu_\mu$ CC events with well-reconstructed muon tracks. Reconstructed neutrino energy and zenith angle histograms from each sample are used in the fit. The flavor composition is parameterized by the electron and tau neutrino fractions at Earth, $(f_{e}, f_{\tau})$, with the muon neutrino fraction given by $f_{\mu} = 1 - f_{e} - f_{\tau}$. These two parameters are included as free variables in the likelihood fit. In addition, the fit includes nuisance parameters accounting for uncertainties in the astrophysical neutrino flux, including its normalization and spectral index, as well as detector systematics such as DOM efficiency and ice properties, following the same configuration used in the flux measurement as described in Section~\ref{sec2}.
\section{Results}\label{sec4}

\subsection{Astrophysical Flux Measurement}
\label{subsec:astrophysical flux measurement}
Detailed results are presented in Ref.~ \cite{zelong_thesis}.  
For a single power law hypothesis the per-flavor astrophysical neutrino flux is parameterized as
\begin{equation}
\Phi^{\nu+\bar{\nu}, per flavor}_{Astro}(E_\nu)=\Phi_{0}\times10^{-18}
\left(\frac{E_\nu}{100~\mathrm{TeV}}\right)^{-\gamma_{Astro}}
\;\mathrm{GeV}^{-1}\,\mathrm{cm}^{-2}\,\mathrm{s}^{-1}\,\mathrm{sr}^{-1},
\end{equation}
where $\Phi_{0}$ is the normalization and $\gamma$ the spectral index.  
The best-fit values are
\begin{equation}
\Phi_{0}=1.83\pm0.21, \qquad \gamma_{Astro}=2.58\pm0.06 .
\end{equation}

\begin{figure}
    \centering
    \begin{subfigure}[b]{0.4\linewidth}
        \includegraphics[width=\linewidth]{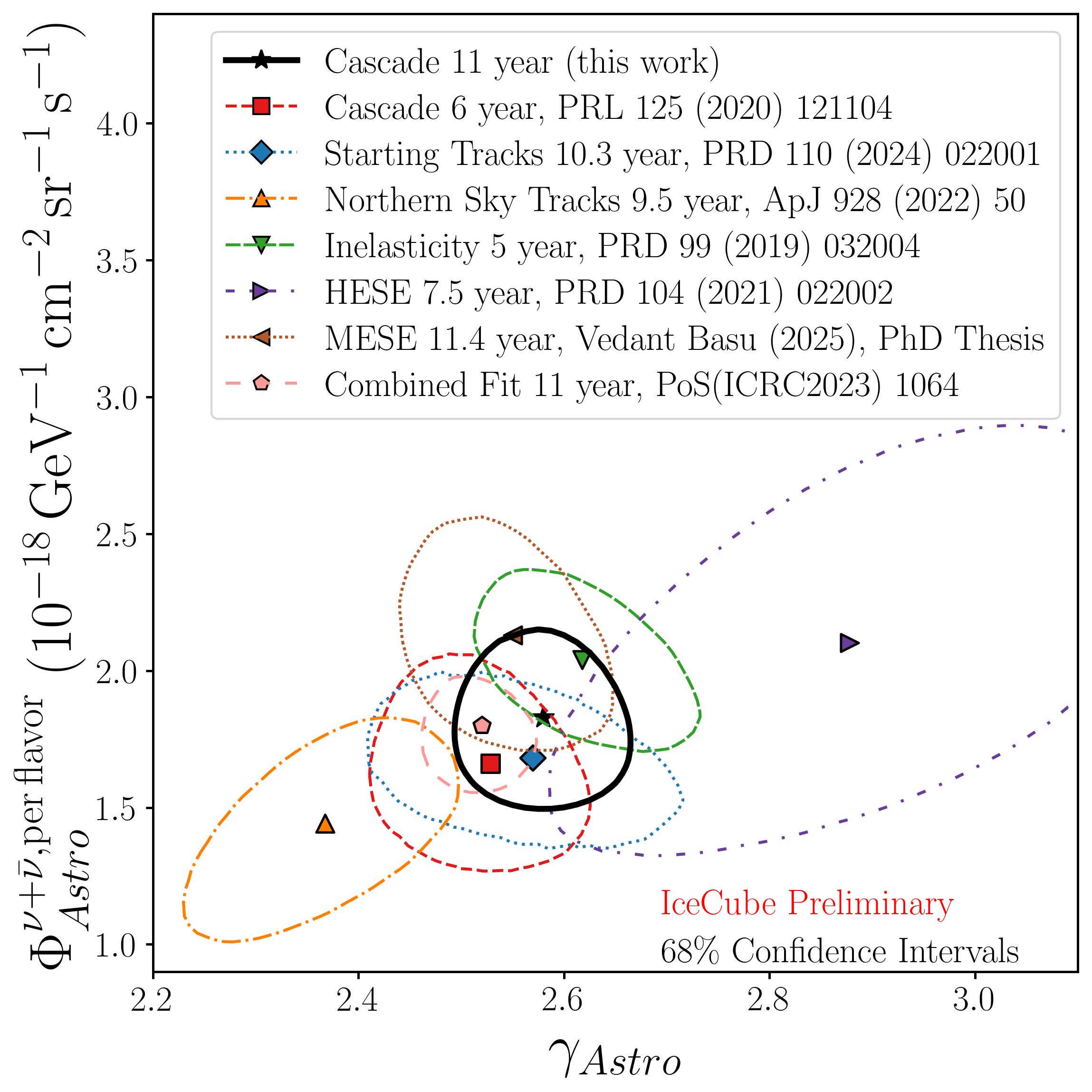}
    \end{subfigure}
    \hfill
    \vspace{-10pt}
    \begin{subfigure}[b]{0.4\linewidth}
        \includegraphics[width=\linewidth]{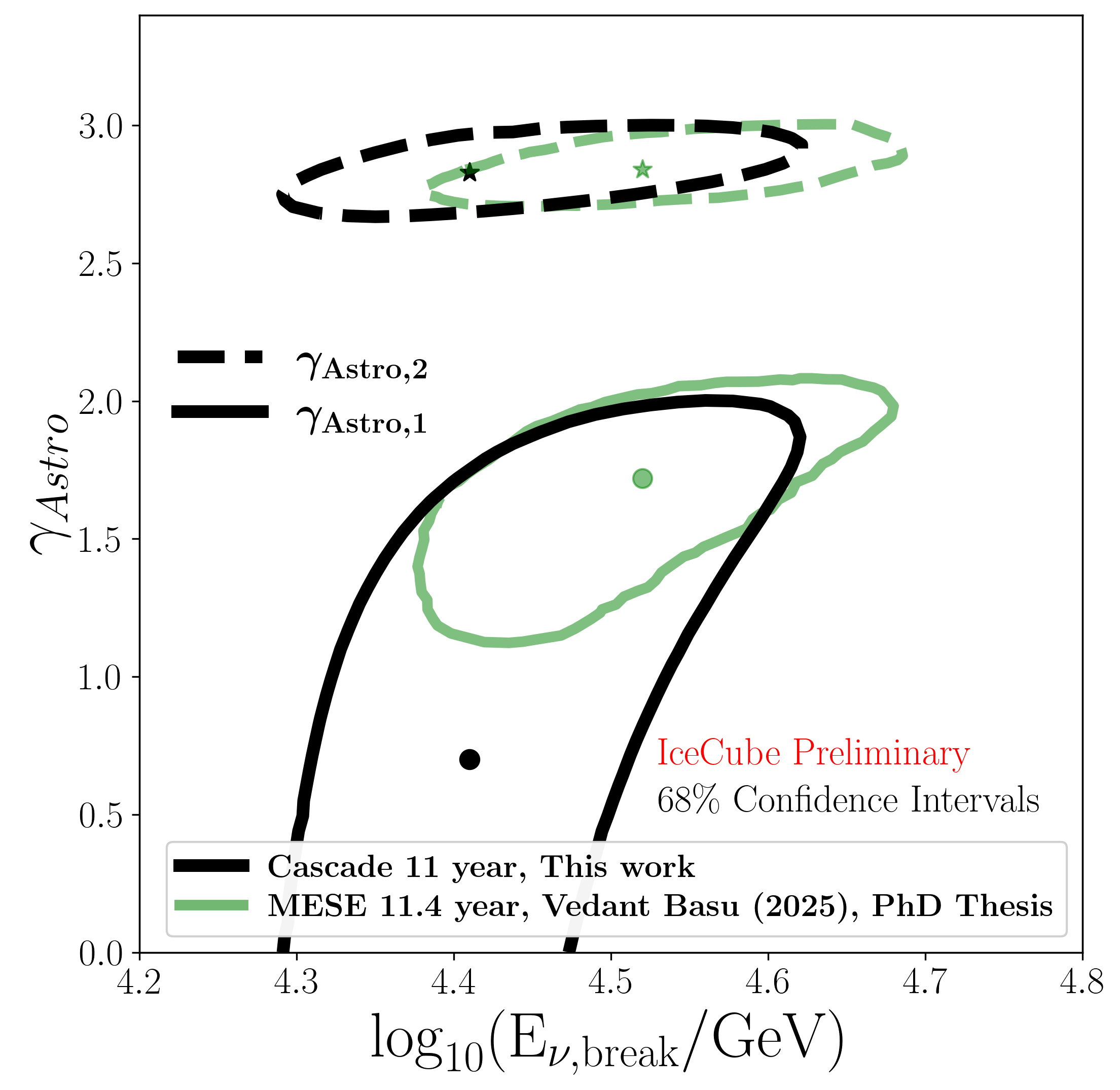}

    \end{subfigure}
    \caption{Comparison of best‐fit contours for single (left) and broken power‐law (right) models with results from other IceCube channels.}
    \label{fig:flux-contours}
\end{figure}

Figure~\ref{fig:flux-contours} compares our result with other IceCube measurements: the previous Cascades \cite{PhysRevLett.125.121104}, the Combined Fit \cite{Naab:2023xcz}, HESE \cite{PhysRevD.104.022002}, MESE \cite{vedant_thesis}, northern-sky track \cite{Abbasi_2022}, starting track \cite{PhysRevD.91.022001}, and the inelasticity analysis \cite{PhysRevD.99.032004}. Our measurement is compatible with all channels. It agrees with the earlier Cascade result within $1\sigma$ confidence interval and benefits from smaller uncertainties thanks to nearly twice the statistics and an improved treatment of systematics. Because this analysis shares an identical astrophysical neutrino single cascade experiments data sample with the Combined Fit while using different atmospheric-muon control samples, simulation sets, and fitting framework, the close overlap of their contours in Fig.~\ref{fig:flux-contours} confirms that the two results are fully consistent and therefore serve as strong mutual validations. The smaller contour size reported in \cite{Naab:2023xcz} stems from a tighter constraint on the atmospheric-muon background achieved by incorporating the northern-sky track sample.

Motivated by the use of a broken power law model \cite{vedant_thesis}, and by the data excess around \(30~\mathrm{TeV}\) \cite{zelong_thesis} that appears under a single power law fit, we also fit a broken power law spectrum: 
\begin{equation}
\Phi_{x}(E_\nu)=\Phi_{b}\times10^{-18}
\begin{cases}
\left(E_\nu/E_{\nu, \textrm{break}}\right)^{-\gamma_{Astro, 1}}, & E_\nu\le E_{\nu, \textrm{break}},\\[4pt]
\left(E_\nu/E_{\nu, \textrm{break}}\right)^{-\gamma_{Astor, 2}}, & E_\nu> E_{\nu, \textrm{break}},
\end{cases}
\;\mathrm{GeV}^{-1}\,\mathrm{cm}^{-2}\,\mathrm{s}^{-1}\,\mathrm{sr}^{-1},
\end{equation}
with best-fit parameters
\begin{equation}
\Phi_{b}=1.72^{+0.37}_{-0.27},\,
\log_{10}\!\left(E_{\nu, \textrm{break}}/\mathrm{GeV}\right)=4.41^{+0.13}_{-0.08},\,
\gamma_{Astro, 1}=0.70^{+1.05}_{-0.70},\,
\gamma_{Astro, 2}=2.83\pm0.12 .
\end{equation}

Figure~\ref{fig:flux-contours} juxtaposes our contours with the MESE result obtained under the broken power law spectral assumption and shows that the two measurements are consistent. MESE constrains the low-energy spectral index more tightly but produces a broader contour for a single-power-law model because the two analyses have different sensitive energy ranges \cite{vedant_thesis, zelong_thesis}. As illustrated at the bottom panel of Figure~\ref{fig:data-MC}, the broken power law provides a better description of the data, especially around $30~\mathrm{TeV}$, as compared to the single power law. A binned \(\chi^{2}\) goodness-of-fit test yields \(\chi^{2}_{\text{SPL}} = 119.7\) for the single power law and \(\chi^{2}_{\text{BPL}} = 97.0\) for the broken power law, underscoring the latter’s superior fit; a detailed discussion of this preference and of the \(30~\mathrm{TeV}\) excess can be found in Ref.~ \cite{zelong_thesis}.

\subsection{Flavor Composition Sensitivity}

\begin{figure}[t]      
  \noindent            
  \begin{minipage}[t]{0.48\linewidth}
    \centering
    \includegraphics[width=\linewidth]{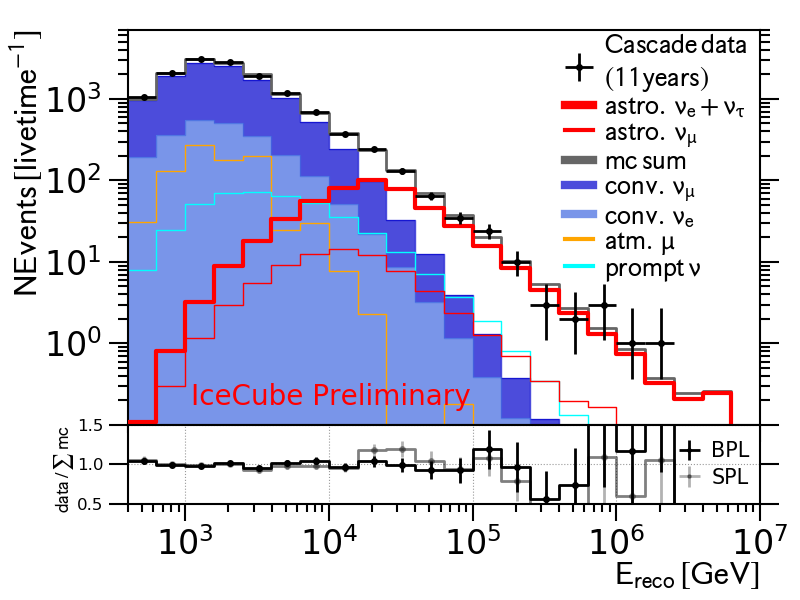}
    \vspace{-25pt}
    \captionof{figure}{Top: Energy distribution of the single cascade 11 year data and MC flux components, assuming BPL model for astrophysical neutrinos. Bottom: Data/MC ratio for SPL and BPL models, highlighting the improvement around $30\,\mathrm{TeV}$.}
    \label{fig:data-MC}
  \end{minipage}\hfill
  \begin{minipage}[t]{0.48\linewidth}
    \centering
    \includegraphics[width=\linewidth]{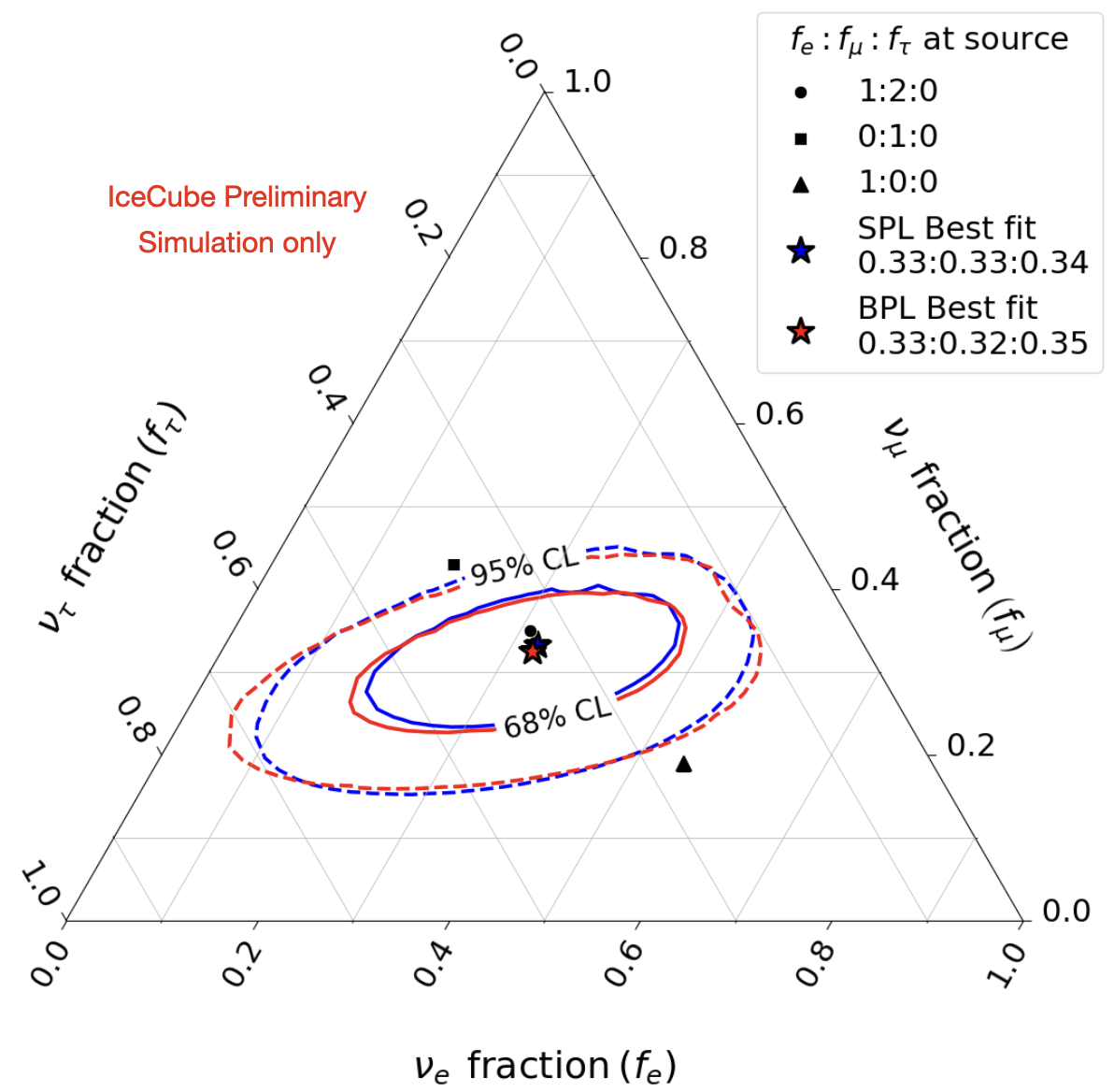}
    \vspace{-25pt}
    \captionof{figure}{Projected 11 years sensitivity to astrophysical neutrino flavor composition at Earth, shown as 68\% and 95\% confidence contours for SPL and BPL models. Black markers denote expected flavor compositions at Earth for common source scenarios.}
    \label{fig:flavor-sensitivity}
  \end{minipage}
  \vspace{-20pt}
\end{figure}

The sensitivity of the combined dataset to the flavor composition is evaluated using an Asimov dataset constructed under the standard oscillation scenario with $(f_{e} : f_{\mu} : f_{\tau}) = (1:1:1)$ at earth using parameters calculated from  \cite{Esteban_2019}. The resulting confidence regions in flavor space are visualized using a flavor triangle plot in Figure~\ref{fig:flavor-sensitivity}, with contours corresponding to 68\% and 95\% confidence levels. The inclusion of our double-cascade $\nu_\tau$ sample enhances sensitivity to the $\nu_\tau$ fraction beyond that achieved by analyses using only cascade and track samples \cite{Lad:2023u9}, and reaches a comparable precision to previous double-cascade-based studies \cite{Lad:2023u9,Basu:2023i3}. 

\section{Conclusion}\label{sec5}

We report two complimentary studies of the diffuse astrophysical neutrino flux in the cascade channel. The most up-to-date measurement results of the diffuse astrophysical neutrino flux, based on 11 years of IceCube data, is reported, which are consistent with those obtained by other IceCube analyses. Data favors a broken power law spectrum over a single power law, a preference driven largely by the 30 TeV excess that emerges under the single power law hypothesis.

In addition, this work demonstrates the capability of IceCube’s latest reconstruction and simulation tools—incorporating updated ice modeling—to enable flavor-sensitive astrophysical neutrino analyses. By identifying well-reconstructed double cascade $\nu_\tau$ events with high purity and accurate decay length reconstruction, we achieve a strong signal-to-background ratio and high-quality input for flavor composition studies. Combining this tau-enriched sample with cascade and track datasets, we provide projected sensitivity to the astrophysical neutrino flavor composition at Earth. The results show improved constraints on the $\nu_\tau$ fraction compared to analyses using only cascade and track samples. This analysis framework strengthens the foundation for future precision measurements of neutrino flavor with IceCube.

\bibliographystyle{ICRC}
\bibliography{references}

\clearpage

\input{authorlist_IceCube.tex}

\end{document}

%% file: ICRCdetails.tex
\ConferenceLogo{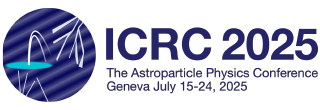}

\FullConference{39th International Cosmic Ray Conference (ICRC2025)\\
 15–24 July 2025\\
Geneva, Switzerland\\}

%% file: authorlist_IceCube.tex
\section*{Full Author List: IceCube Collaboration}

\scriptsize
\noindent
R. Abbasi$^{16}$,
M. Ackermann$^{63}$,
J. Adams$^{17}$,
S. K. Agarwalla$^{39,\: {\rm a}}$,
J. A. Aguilar$^{10}$,
M. Ahlers$^{21}$,
J.M. Alameddine$^{22}$,
S. Ali$^{35}$,
N. M. Amin$^{43}$,
K. Andeen$^{41}$
% ...
\\
\\
% ...
$^{10}$ Universit{\'e} Libre de Bruxelles, Science Faculty CP230, B-1050 Brussels, Belgium \\
% ...
$^{16}$ Department of Physics, Loyola University Chicago, Chicago, IL 60660, USA \\
$^{17}$ Dept. of Physics and Astronomy, University of Canterbury, Private Bag 4800, Christchurch, New Zealand \\
% ...
$^{21}$ Niels Bohr Institute, University of Copenhagen, DK-2100 Copenhagen, Denmark \\
$^{22}$ Dept. of Physics, TU Dortmund University, D-44221 Dortmund, Germany \\
% ...
$^{35}$ Dept. of Physics and Astronomy, University of Kansas, Lawrence, KS 66045, USA \\
% ...
$^{39}$ Dept. of Physics and Wisconsin IceCube Particle Astrophysics Center, University of Wisconsin{\textemdash}Madison, Madison, WI 53706, USA \\
% ...
$^{41}$ Department of Physics, Marquette University, Milwaukee, WI 53201, USA \\
% ...
$^{43}$ Bartol Research Institute and Dept. of Physics and Astronomy, University of Delaware, Newark, DE 19716, USA \\
% ...
$^{63}$ Deutsches Elektronen-Synchrotron DESY, Platanenallee 6, D-15738 Zeuthen, Germany \\
$^{\rm a}$ also at Institute of Physics, Sachivalaya Marg, Sainik School Post, Bhubaneswar 751005, India \\
% ...

\subsection*{Acknowledgments}

\noindent
The authors gratefully acknowledge the support from the following agencies and institutions:
USA {\textendash} U.S. National Science Foundation-Office of Polar Programs,
U.S. National Science Foundation-Physics Division,
U.S. National Science Foundation-EPSCoR,
U.S. National Science Foundation-Office of Advanced Cyberinfrastructure,
Wisconsin Alumni Research Foundation,
Center for High Throughput Computing (CHTC) at the University of Wisconsin{\textendash}Madison,
Open Science Grid (OSG),
Partnership to Advance Throughput Computing (PATh),
Advanced Cyberinfrastructure Coordination Ecosystem: Services {\&} Support (ACCESS),
Frontera and Ranch computing project at the Texas Advanced Computing Center,
U.S. Department of Energy-National Energy Research Scientific Computing Center,
Particle astrophysics research computing center at the University of Maryland,
Institute for Cyber-Enabled Research at Michigan State University,
Astroparticle physics computational facility at Marquette University,
NVIDIA Corporation,
and Google Cloud Platform;
Belgium {\textendash} Funds for Scientific Research (FRS-FNRS and FWO),
FWO Odysseus and Big Science programmes,
and Belgian Federal Science Policy Office (Belspo);
Germany {\textendash} Bundesministerium f{\"u}r Forschung, Technologie und Raumfahrt (BMFTR),
Deutsche Forschungsgemeinschaft (DFG),
Helmholtz Alliance for Astroparticle Physics (HAP),
Initiative and Networking Fund of the Helmholtz Association,
Deutsches Elektronen Synchrotron (DESY),
and High Performance Computing cluster of the RWTH Aachen;
Sweden {\textendash} Swedish Research Council,
Swedish Polar Research Secretariat,
Swedish National Infrastructure for Computing (SNIC),
and Knut and Alice Wallenberg Foundation;
European Union {\textendash} EGI Advanced Computing for research;
Australia {\textendash} Australian Research Council;
Canada {\textendash} Natural Sciences and Engineering Research Council of Canada,
Calcul Qu{\'e}bec, Compute Ontario, Canada Foundation for Innovation, WestGrid, and Digital Research Alliance of Canada;
Denmark {\textendash} Villum Fonden, Carlsberg Foundation, and European Commission;
New Zealand {\textendash} Marsden Fund;
Japan {\textendash} Japan Society for Promotion of Science (JSPS)
and Institute for Global Prominent Research (IGPR) of Chiba University;
Korea {\textendash} National Research Foundation of Korea (NRF);
Switzerland {\textendash} Swiss National Science Foundation (SNSF).